\documentclass[11pt,preprint]{aastex}
\pdfoutput=1
\usepackage{graphicx}

\let\oldfootsep=\footnotesep
\newcommand\ltsima{$\; \buildrel <\over\sim \;$}
\newcommand\simlt{\lower.5ex\hbox{\ltsima}}
\newcommand\gtsima{$\; \buildrel >\over\sim \;$}
\newcommand\simgt{\lower.5ex\hbox{\gtsima}}

\newcommand\msun {M_\odot}

\newcommand\mearth {{M_\oplus}}

\newcommand\ie{{i.e.}}

\newcommand{\mathbold}[1]{\mbox{\boldmath $\bf#1$}}

\newcommand\mubold{{\mathbold \mu}}

\newcommand\vbold{{\mathbold v}}
\shorttitle{}
\shortauthors{Bennett et al}

\begin{document}


\title{Confirmation of the Planetary Microlensing Signal and
Star and Planet Mass Determinations for Event OGLE-2005-BLG-169}


\author{D.P.~Bennett\altaffilmark{1},
A.~Bhattacharya\altaffilmark{1},
J.~Anderson\altaffilmark{2},
I.A.~Bond\altaffilmark{3},
N.~Anderson\altaffilmark{4}, 
R.~Barry\altaffilmark{4}, 
V.~Batista\altaffilmark{5},
J.-P.~Beaulieu\altaffilmark{5},
D.L.~DePoy\altaffilmark{6},
Subo~Dong\altaffilmark{7},
B.S.~Gaudi\altaffilmark{8},
E.~Gilbert\altaffilmark{4},  
A.~Gould\altaffilmark{8},
R.~Pfeifle\altaffilmark{4},  
R.W.~Pogge\altaffilmark{8}, 
D.~Suzuki\altaffilmark{1}, 
S.~Terry\altaffilmark{4},  and
 A.~Udalski\altaffilmark{9}
             } 
              

\altaffiltext{1}{Department of Physics,
    University of Notre Dame, 225 Nieuwland Science Hall, Notre Dame, IN 46556, USA; 
    Email: {\tt bennett@nd.edu}}
\altaffiltext{2}{Space Telescope Science Institute, 3700 San Martin Drive, Baltimore, MD 21218, USA}
\altaffiltext{3}{Institute of Natural and Mathematical Sciences, Massey University, Auckland 0745, New Zealand}
\altaffiltext{4}{Laboratory for Exoplanets and Stellar Astrophysics, NASA/Goddard Space Flight Center, Greenbelt, MD 20815, USA}
\altaffiltext{5}{UPMC-CNRS, UMR 7095, Institut d'Astrophysique de Paris, 98 bis boulevard Arago, F-75014 Paris, France}
\altaffiltext{6}{Department of Physics, Texas A\&M University, College Station, TX 77843-4242, USA}
\altaffiltext{7}{Kavli Institute for Astronomy and Astrophysics, Peking University, Hai Dian District, Beijing 100871, China}
\altaffiltext{8}{Department of Astronomy, Ohio State University, 140 West 18th Avenue, Columbus, OH 43210, USA}
\altaffiltext{9}
{Warsaw University Observatory, Al.~Ujazdowskie~4, 00-478~Warszawa,Poland}



\begin{abstract}
We present {\it Hubble Space Telescope} (HST) Wide Field Camera 3 (WFC3)
observations of the source and lens stars for planetary microlensing event
OGLE-2005-BLG-169, which confirm the relative proper motion prediction
due to the planetary light curve signal observed for this event. This (and the
companion Keck result) provide the first confirmation of a planetary 
microlensing signal, for which the deviation was
only 2\%. The follow-up observations determine the flux of the planetary host
star in multiple passbands and remove light curve model ambiguity
caused by sparse sampling of part of the light curve. This leads to a precise
determination of the properties of the OGLE-2005-BLG-169Lb planetary
system. Combining the constraints from the microlensing light curve 
with the photometry and astrometry of the HST/WFC3 data, we find star
and planet masses of $M_* = 0.69 \pm 0.02\msun$ and $m_p = 14.1\pm 0.9\mearth$.
The planetary microlens system is located toward the Galactic bulge
at a distance of $D_L = 4.1\pm 0.4\,$kpc
and the projected star-planet separation is $a_\perp = 3.5\pm 0.3\,$AU, corresponding
to a semi-major axis of $a = 4.0{+2.2\atop -0.6}\,$AU.
\end{abstract}


\section{Introduction}
\label{sec-intro}

Gravitational microlensing is unique among planet
detection methods \citep{bennett_rev,gaudi_araa} in its sensitivity to planets
with masses smaller than Earth \citep{bennett96} orbiting beyond the snow
line \citep{mao91,gouldloeb92}, where planet formation is thought to be the most 
efficient \citep{ida05,lecar_snowline,kennedy-searth,kennedy_snowline,thommes08},
according to the core accretion theory of planet formation \citep{lissauer_araa,pollack96}.
Microlensing is also able to detect planets orbiting stars at distances ranging from
a few hundred parsecs up to $D_L \simeq 8\,$kpc. Since the microlensing method
doesn't depend on light from the planetary host star, it can be used to find planets
orbiting very faint star or even stellar remnants or brown dwarfs \citep{bennett_rev,gaudi_araa}. 
However, one
drawback of the microlensing method is that the microlensing light curves usually do not
indicate the planet or host star mass. Instead, they generally yield the planet-star
mass ratio, $q$, and the separation in units of the Einstein radius ($R_E$),
except for events that exhibit the microlensing parallax effect 
\citep{gaudi-ogle109,bennett08,muraki11,moa328}. A measurement of the microlensing 
parallax effect for a planetary microlensing event usually provides enough information
about the lensing geometry to determine the lens mass. The mass measurement
does require that the angular Einstein radius, $\theta_E = R_E/D_L$, be known, but this can
be determined for most planetary events from finite source effects in the light curve
that allow the source radius crossing time, $t_*$, to be  measured.
However, most events do not
have a measurable microlensing parallax effect, particularly those due to lens
systems in the Galactic bulge.

A more generally applicable method to determine the lens system mass is to
detect the host star, for an event in which $\theta_E$ has been determined.
This requires high angular resolution imaging because the lens and
source stars are not resolved from unrelated stars in ground-based, 
seeing-limited images. 
When $\theta_E$ is known, it provides a mass-distance relation for the lens
system, and this can be combined with a mass-luminosity relation to
determine the mass of the lens system. This has been done for a number
of events \citep{bennett06,dong-ogle71,moa192_naco,batista14}, but sometimes it isn't
clear if the excess flux is really due to the lens star \citep{sumi10,janczak10,gould-1dpar},
as unrelated stars or companions to the source or lens star cannot always
be excluded. The keys to establishing that the excess flux is due to the
planetary host (and lens) star are to measure lens brightness in multiple 
pass bands and to measure the relative lens-source proper, $\mu_{\rm rel}$,
which is usually known from the light curve.

In this paper, we present the first direct measurements of the relative
proper motion, $\mu_{\rm rel}$, for a planetary microlensing event,
OGLE-2005-BLG-169, using HST observations in three 
Wide Field Camera 3 (WFC3) passbands: F814W, F555W, and F438W.
The light curve prediction of $\mu_{\rm rel}$ comes from the planetary
signal itself, so our confirmation of this prediction is a confirmation
of the planetary signal. Thus, the planetary signal for
OGLE-2005-BLG-169Lb is the first to be confirmed by follow-up 
observations. The HST follow-up observations also provide a tighter
constraint on $\mu_{\rm rel}$ than the light curve does,
so we are able to obtain tighter constraints on the light curve
parameters than the discovery paper \citep{gould06}.
The HST lens brightness measurements, when combined
with the $\theta_E$ mass-distance relation, yield the masses
and distance of the planet and its host star, as well as their projected 
separation. A companion paper \citep{batista15} presents independent 
measurements of $\mu_{\rm rel}$ and the lens brightness in the $H$-band 
using adaptive optics observations from the {\it Keck-II Telescope}. These
{\it Keck} measurements are consistent with the HST results presented
here.

This paper is organized as follows. We discuss the light curve data 
and photometry in Section~\ref{sec-lc_data}, and in Section~\ref{sec-lc}
we present the light curve models that are consistent with the data. In
Section~\ref{sec-radius}, we show how the angular radius of the source
star relates to its color and brightness. Then in
Section~\ref{sec-hst_data}, we describe the HST data and its
reduction, and in Section~\ref{sec-conf} we compare the lens-source
relative proper motion prediction from the light curve with the
HST measurement. In Section~\ref{sec-Keck} we compare our results to
the Keck adaptive optics observations made 1.74 years later and show that
the combined HST and Keck observations confirm that our identification of 
the lens star is correct. The constraints on the lens system from the HST data 
are explored in Section~\ref{sec-const}.
Finally in Section~\ref{sec-conclude}, we present our
conclusions and explain how this analysis demonstrates the
primary exoplanet host mass measurement method for the
WFIRST and EUCLID missions \citep{bennett02,bennett07,WFIRST_rep,WFIRST_AFTA,penny13}.

\section{Light Curve Data and Photometry}
\label{sec-lc_data}

OGLE-2005-BLG-169 is unique among planetary microlensing events
in a number of respects \citep{gould06}. It has the smallest impact parameter, $u_0$,
of any planetary microlensing event, and it has the smallest amplitude 
photometric signal of any planetary microlensing event. The planetary
signal entirely in the extremely high cadence data taken from the $2.4\,$m 
MDM telescope. (More than 1000 observations were taken in a 3-hour period
at high magnification.) Because of the low amplitude signal, there was concern 
that the data could be contaminated by systematic photometry errors. Due to 
this concern, the MDM data were reduced with two independent photometry 
pipelines, the OGLE pipeline \citep{ogle-pipeline} and the \citet{hartmanISIS}
implementation of the \citet{ala98} photometry code. This later reduction was
performed by K.Z.\ Stanek, and we will refer to it as the Stanek reduction.

\begin{figure}
\plotone{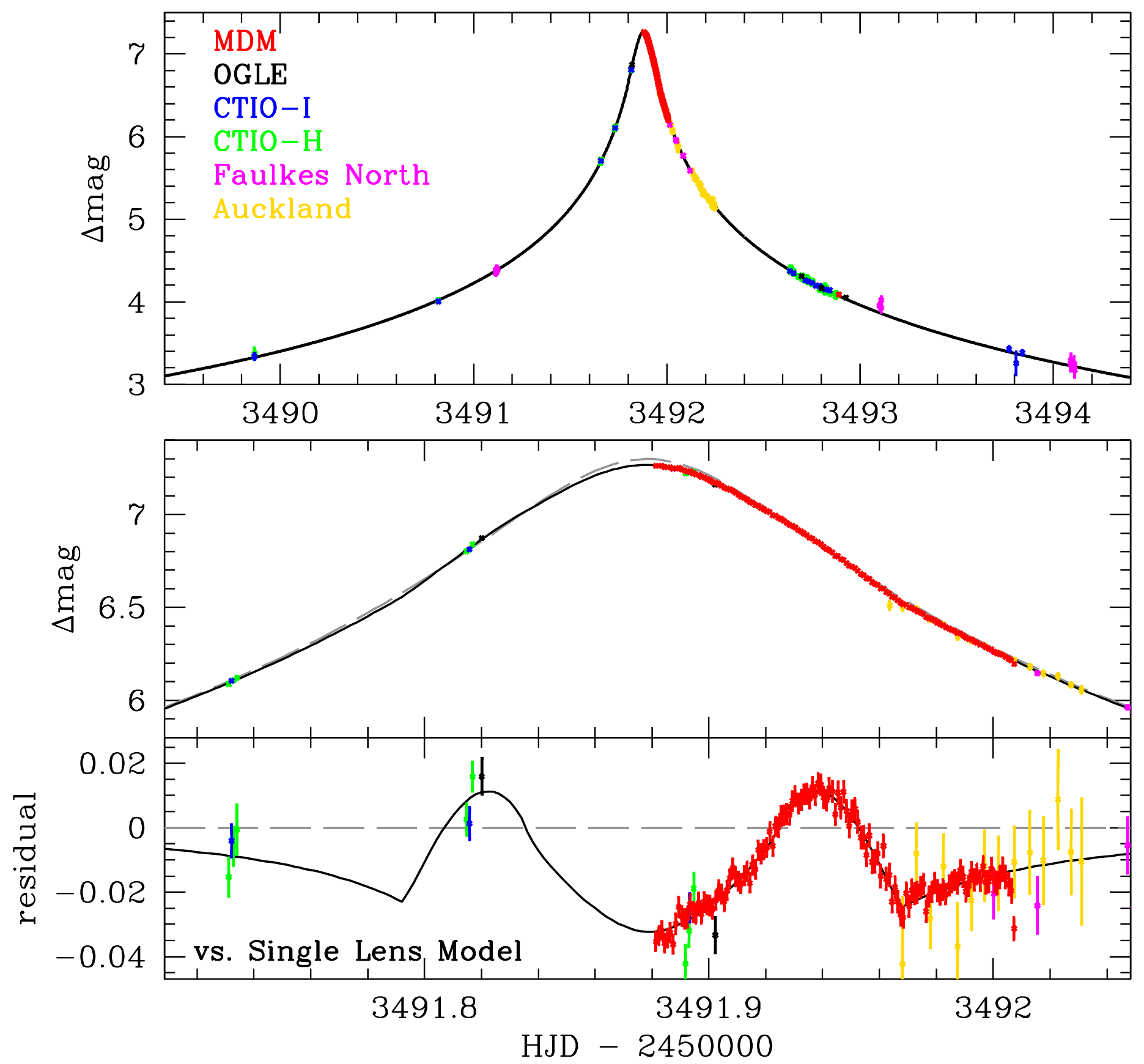}
\caption{The light curve peak of event OGLE-2005-BLG-169 with photometric
measurements from the MDM 2.4m $I$-band (red, Stanek reduction), 
OGLE $I$-band (black), CTIO $I$ and $H$-bands (blue
and green), Faulkes Telescope North (magenta), and the Auckland unfiltered 
telescope (gold). The best fit model is indicated by the black curve,
and the grey dashed curve indicates the same model without the planetary
signal. The bottom panel shows the residual with respect to this no-planet
model. The MDM data clearly trace out the caustic exit feature, but
the data on the rising side provide a very weak constraint on the caustic
entry properties. So, a variety of angles between the lens axis
and the source trajectory are permitted by the photometry. The best
fit model to the data set including the Stanek MDM photometry presented
here is consistent with our proper motion measurement, while the light
curve presented in \citet{gould06} is not.
\label{fig-lc}}
\end{figure}

In addition to the MDM data set, which contains the planetary signal, the 
photometry for this event include data from the $1.3\,$m OGLE survey
telescope (responsible for the identification of the microlensing event), 
the $1.3\,$m SMARTS telescope at CTIO in Chile, the $2.0\,$m Faulkes Telescope 
North in Hawaii, and the $0.35\,$m Nustrini Telescope in Auckland, 
New Zealand. We use the same photometric reduction for each data
set that was used by \citet{gould06} except for the CTIO data. A minor problem
was discovered in the CTIO $I$-band photometry used in the original paper.
The CTIO $I$-band photometry yielded a source magnitude
that was $0.13\,$mag fainter than the source magnitude from the OGLE $I$-band 
photometry when the photometry from both data sets was calibrated to the
OGLE-III photometry database \citep{ogle3-phot}. This inconsistency was largely
resolved (reduced to $0.03\,$mag) by switching from DoPHOT \citep{dophot} to SoDoPHOT 
photometry \citep{bennett-sod}. In this analysis, we have also included
the CTIO $H$-band data, taken simultaneously with the $V$ and $I$-band
data on the Andicam instrument on the SMARTS telescope. The $H$-band data
is especially useful because they allow a more precise determination of the 
angular source radius \citep{kervella_dwarf,boyajian14}. The $H$-band light
curve used in this paper is a SoDoPHOT reduction, but a reduction using
the MOA Collaboration difference imaging pipeline \citep{bond01} gives
indistinguishable results.

\section{Light Curve Models}
\label{sec-lc}

The light curve models used for this paper are different from the models
presented in \citet{gould06} because a different data set is used. We use
the \citet{bennett-himag} modeling code instead of the \citet{gould06} code, but this
has no effect on the results, as these codes have been shown to give
identical results to better than 1 part in $10^4$. Our conclusions based on the 
light curve modeling alone are essentially the same as the conclusions of 
\citet{gould06} . As discussed in \citet{gould06} and Section~\ref{sec-lc_data}, the 
planetary signal for this event is particularly sensitive to potential systematic 
photometry errors because of the larger than usual S/N of the MDM observations 
and the small amplitude of the planetary signal. For this reason, \citet{gould06} 
did the complete analysis using both the Stanek and OGLE-pipeline photometry.
We continue this philosophy in this paper and
assume that the OGLE-pipeline and Stanek reductions are equally likely
to be correct, and so we perform Markov Chain Monte Carlo (MCMC) calculations
for both data sets starting at the parameters of each of the local 
$\chi^2$ minima presented in \citet{gould06}.

The results of these MCMC calculations differ in detail from the 
results presented in \citet{gould06}, in the sense that the
models with the source trajectory nearly perpendicular to the lens
axis are now somewhat favored with respect to the previous
analysis. With the Stanek version of the MDM photometry, these
models are now favored by $\Delta\chi^2 = 8.8$ over the best
fit model with a source trajectory $>25^\circ$ from perpendicular
to the lens axis. The best fit model using the Stanek version of the
MDM photometry is presented in Figure~\ref{fig-lc}, and the parameters
of this model and the best fit $s<1$ model are given in Table~\ref{tab-mparams}. 
The best fit model, which has $s>1$, is labeled as ``Stanek $s>1$\rlap,"
and the parameters of the best fit $s<1$ are also given. This $s<1$ 
model is very slightly disfavored with $\Delta\chi^2 = 0.12$. Table~\ref{tab-mparams}
also gives the MCMC averages of the parameters both without the
constraints from the HST measurements (in the next-to-last column)
and with the constraints from the HST measurements in the last column.
Because of the wide variation in the  $\theta$ values (source trajectory angles)
allowed by the light curve, there is a large scatter in some of the other fit
parameters, such as the source radius crossing time, $t_*$, and the planet:star
mass ratio, $q$.

\begin{deluxetable}{cccccc}
\tablecaption{Model Parameters
                         \label{tab-mparams} }
\tablewidth{0pt}
\tablehead{
& & & & \multicolumn{2}{c} {MCMC averages} \\
\colhead{parameter}  & \colhead{units} &
\colhead{Stanek $s>1$} & \colhead{Stanek $s<1$} &\colhead{ no const.} & \colhead{$\mubold_{\rm rel,H}$ const.} 
}  

\startdata

$t_E$ & days & 43.09 & 43.16 & 41.8(2.9) & 42.5(1.4) \\
$t_0$ & ${\rm HJD}-2453490$ & 1.8784 & 1.8784 & 1.8776(10) &  1.8784(1) \\
$u_{\rm min}$ & &0.001229 &0.001228 & 0.001267(9) & 0.001250(4)  \\
$s$ & & 1.0190 & 0.9828 & 1.004(18) &  1.001(18) \\
$\theta$ & radians & 1.6025 & 1.6069 & 1.43(20) & 1.60(3)  \\
$q$ & $10^{-5}$ & 5.913 & 5.844 & 7.07(1.22) & 6.15(30)  \\
$t_\ast$ & days & 0.02174 & 0.02168 & 0.0202(17) & 0.0228(5)  \\
$\theta_E$ & mas & 0.905 & 0.911 & 0.965(94) & 0.848(27) \\
$\mu_{\rm rel,G}$ & mas/yr & 7.67 & 7.69 & 8.47(87) & 7.29(15) \\
$H_s$ & & 18.852 & 18.854 & 18.81(8) & 18.84(4)  \\
$I_s$ & & 20.592 & 20.594 & 20.55(8) & 20.58(4)  \\
$V_s$ & & 22.254 & 22.257 & 22.21(8) & 22.24(4)  \\
fit $\chi^2$ &  & 1146.66 & 1146.78 &  &   \\

\enddata
\end{deluxetable}

\begin{figure}
\epsscale{0.7}
\plotone{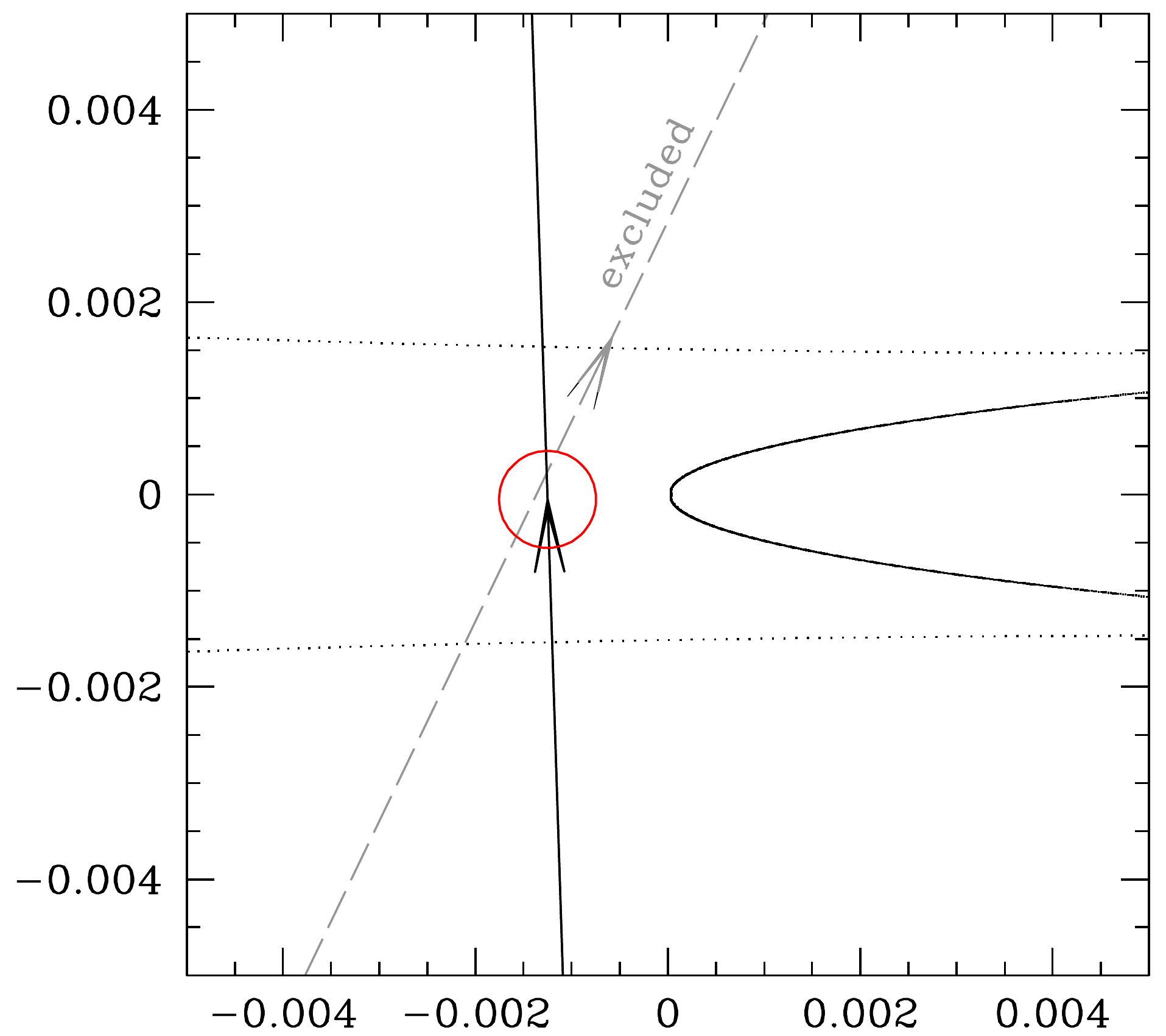}
\caption{The caustic configuration for the OGLE-2005-BLG-169 model
shown in Figure~\ref{fig-lc}. The black line, with arrow, shows the
source trajectory for this model, while the grey dashed line shows
the source trajectory for the other local $\chi^2$ minimum for the 
light curve modeling. This and similar models are consistent with the light curve,
but they are contradicted by the relative proper motion measurement
that we present here.
\label{fig-caustic}}
\end{figure}

Figure~\ref{fig-caustic} shows a close-up of the caustic configuration 
for the best-fit model with the source trajectory given by the solid
black line. The red circle indicates the size of the source star, 
and the gray dashed line shows the source trajectory
for the model presented in  \citet{gould06}.

\section{Calibration and Source Radius}
\label{sec-radius}

\begin{figure}
\epsscale{0.7}
\plotone{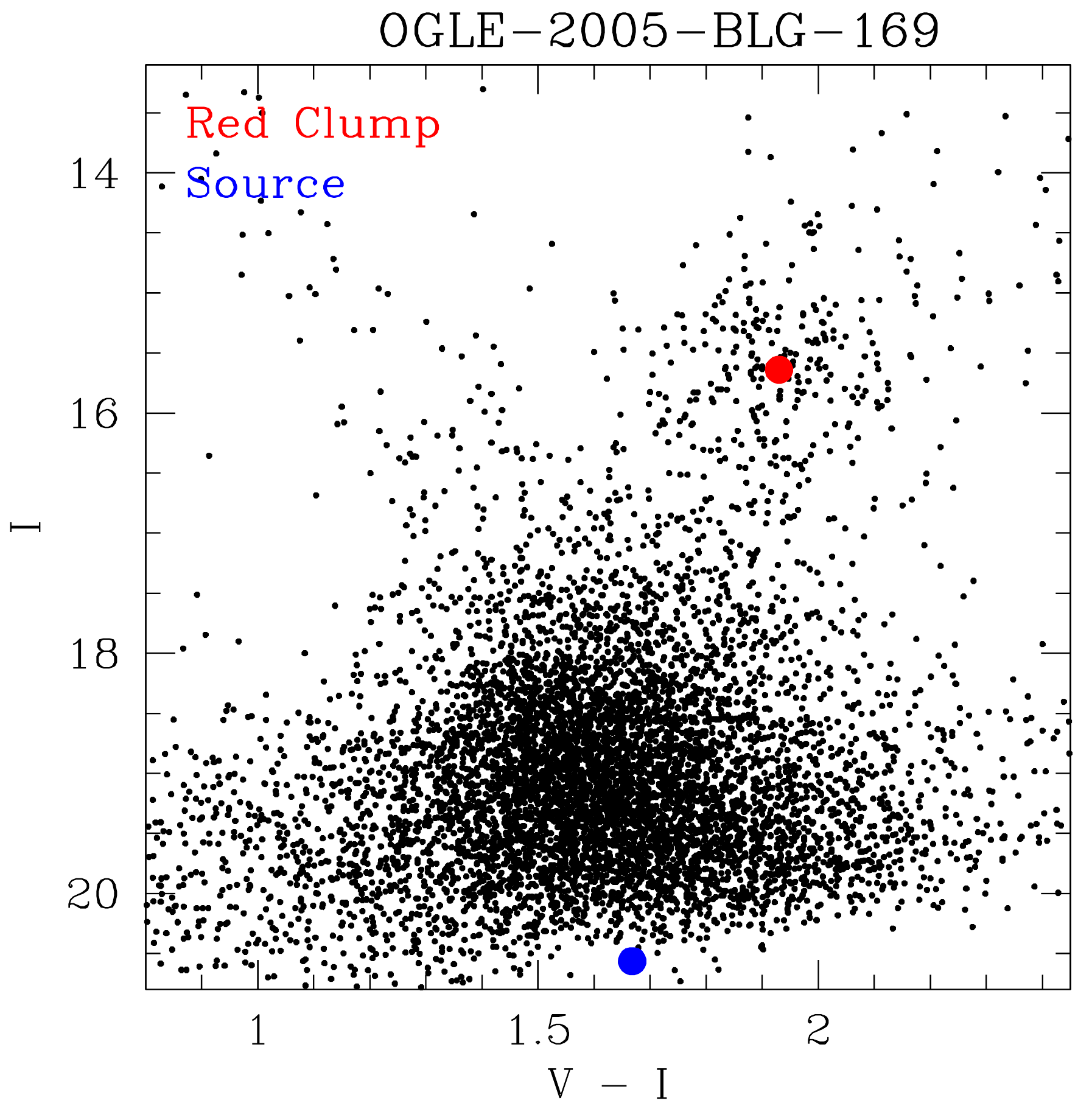}
\caption{The $(V-I,I)$ color magnitude diagram (CMD) of the stars in the OGLE-III catalog
\citep{ogle3-phot}
within $120^{\prime\prime}$ of OGLE-2005-BLG-169. The red spot indicates
red clump giant centroid, and the blue spot indicates the source magnitude
and color.
\label{fig-cmd}}
\end{figure}

%
In order to measure the angular Einstein radius, $\theta_E = \theta_\ast t_E/t_\ast$, 
we must determine the angular radius of the source star, $\theta_\ast$,
from the dereddened brightness and color of the 
source star \citep{kervella_dwarf,boyajian14}. We determine the source star
brightness in the $V$ and $I$-bands by calibrating the CTIO $V$-band 
and OGLE $I$-band magnitudes to the OGLE-III catalog \citep{ogle3-phot} yielding
the following relations:
\begin{eqnarray}
V_{\rm O3cal} &= 23.08516 + 0.97257\, V_{\rm Sod} +  0.02743\, I_{\rm O3lc} \pm 0.004 
\label{eq-V-cal} \\
I_{\rm O3cal} &= 1.406255 + 0.93933\, I_{\rm O3lc} + 0.060674\, V_{\rm Sod}  \pm 0.004 \ .
\label{eq-I-cal}
\end{eqnarray}
$I_{\rm O3lc}$ is the OGLE $I$-band light curve magnitude, which differs from the
standard Cousins $I$-band used in the OGLE-III catalog, while $V_{\rm O3cal}$ and
$I_{\rm O3cal}$ refer to the Johnson $V$-band and Cousins $I$-band magnitudes,
as presented in the OGLE-III catalog \citep{ogle3-phot}. $V_{\rm Sod}$ is the raw
CTIO $V$-band photometry from our SoDoPHOT reduction. The $V$-band calibration
is based on 54 stars with $ 1.0 \leq (V-I)_{\rm O3cal} < 2.2$ and $I_{\rm O3cal} \leq 16.0$
within 2 arc minutes of the target star, and the $I$-band calibration employs
the formulae presented in \citet{ogle3-phot}. 

Our CTIO $H$-band SoDoPHOT magnitudes are calibrated to 2MASS \citep{2mass_cal} 
with the following relation,
\begin{equation}
H_{\rm 2mass} = H_{\rm Sod} + 19.849 \pm 0.010 \ ,
\label{eq-H-cal}
\end{equation}
based on 36 stars within $105^{\prime\prime}$ of the 
target.

In order to estimate the source radius, we need extinction-corrected magnitudes,
and we determine these from the magnitudes and colors of the centroid of the
red clump giant feature in the color magnitude diagram (CMD), as indicated
in Figure~\ref{fig-cmd}. The extinction can be determined most accurately if
three colors are used \citep{bennett-ogle109}, and we find that the
red clump centroid in this field is at $I_{\rm cl} = 15.61$, $(V-I)_{\rm cl} = 1.93$,
$(I-H)_{\rm cl} = 2.07$, which implies $H_{\rm cl} = 13.54$ and $V_{\rm cl} = 17.54$.

We follow the method of \citet{bennett-ogle109} to determine the extinction, but
we use the updated dereddened red clump magnitudes of \citet{nataf13}. We assume
absolute red clump giant centroid magnitudes of $M_{H\rm cl} = -1.30$,
$M_{I\rm cl} = -0.13$, and $M_{V\rm cl} = 0.93$. The
Galactic coordinates of OGLE-2005-BLG-169 are $(l,b) = (0.6769^\circ,  -4.7402^\circ)$,
and this implies a distance modulus of $DM = 14.541$.
Using the \citet{bennett-ogle109} method, we estimate the extinction toward the
center of the Galaxy in this direction to be $A_H = 0.374 \pm 0.020$, 
$A_I = 1.256 \pm 0.050$, and $A_V = 2.132 \pm 0.090$. These extinction
values allow us to determine the dereddened magnitude for each passband,
$C_{s0} = C_s - A_C$ where $C$ refers to the passband (either $V$, $I$, or
$H$).

These dereddened magnitudes can be used to determine the angular source radius,
$\theta_*$. Of the measured source magnitudes, the most precise determination
of $\theta_*$ comes from the $(V-H),H$ relation. We use
\begin{equation}
\log_{10}\left[2\theta_*/(1 {\rm mas})\right] = 0.536654 + 0.072703\,(V-H)_{s0} -0.2\,H_{s0} \ ,
\end{equation}
which comes from the \citet{boyajian14} analysis. These numbers are not included
in the \citet{boyajian14} paper, but they were provided in a private communication from 
T.S.\ Boyajian (2014). She reports that this formula determines $\theta_*$ better than 
2\% accuracy. This is somewhat better than the 2.6\% accuracy of the $(V-H),H$ relation
of \citet{kervella_dwarf}. (They report an accuracy of 1.12\% for $\log_{10}(\theta_*)$,
which corresponds to 2.6\% accuracy for $\theta_*$.)

\begin{figure}
\epsscale{0.6}
\plotone{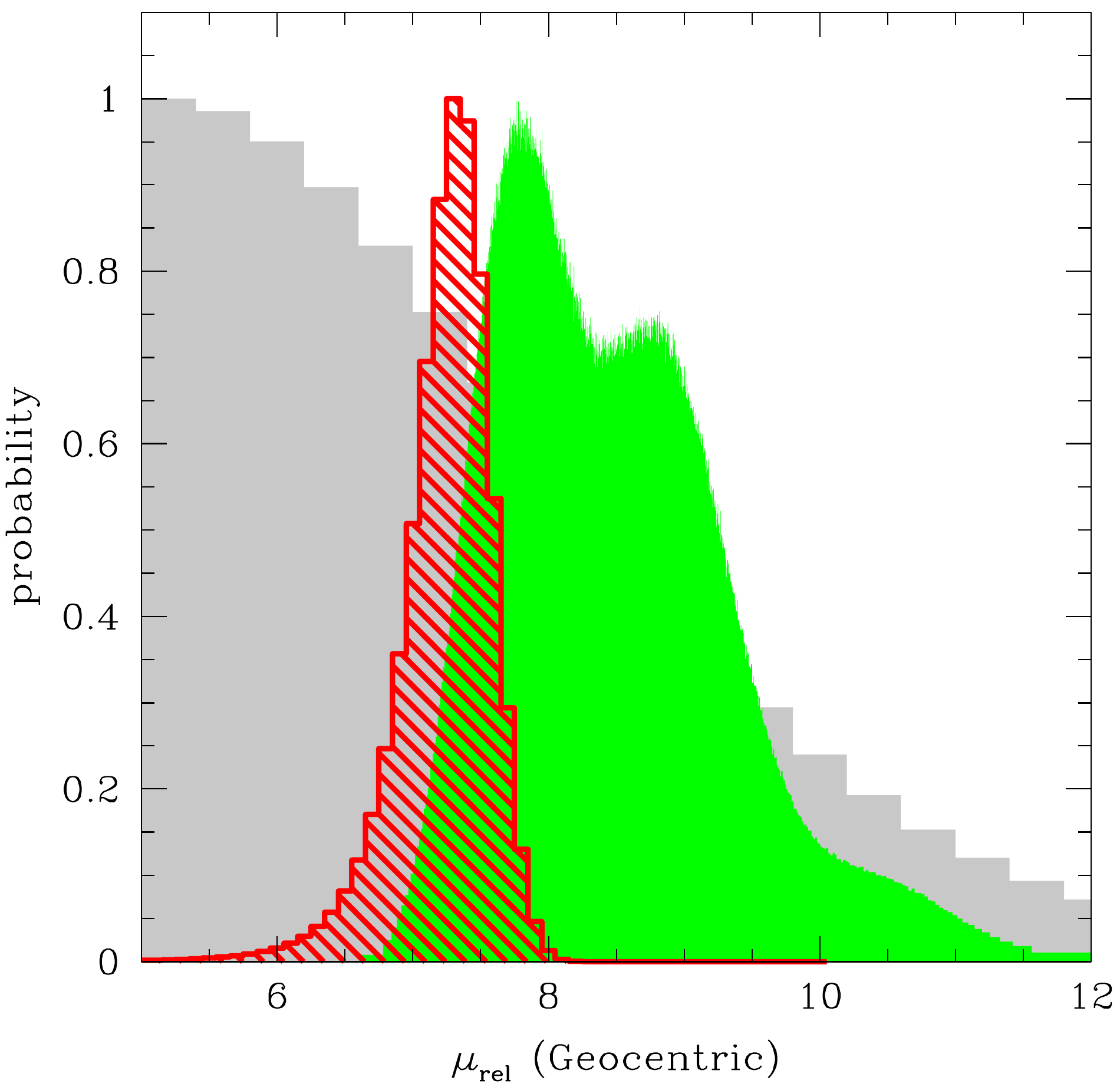}
\caption{Comparison of geocentric proper motion predictions
from all Galactic bulge microlensing events (grey histogram), microlensing
models consistent with the light curve data
(green histogram) and with the HST measurement
(red cross-hatched histogram). The light curve $\mu_{\rm rel,G}$ distribution
is drawn from MCMC calculations using both the Stanek and OGLE MDM
reductions. The HST $\mu_{\rm rel,G}$ distribution has been converted
from the two-dimensional $\mubold_{\rm rel,H}$ measurement using
a probability distribution for relative lens distance ($D_L/D_S$) from
a Galactic model.
\label{fig-pm_comp}}
\end{figure}

The implied source radii for the best fit $s>1$ and $s<1$ models are given in 
Table~\ref{tab-mparams}, along with the angular Einstein radius, $\theta_E$,
and the lens-source relative proper motion, $\mu_{\rm rel,G} = \theta_*/t_*$, in a 
geocentric reference frame. The light curve parameter that the relative
proper motion depends on is the source radius crossing time, $t_*$, which
is measured in the reference frame of the Earth-bound observatories
that observe the light curve. So, $t_*$ is measured the Geocentric reference
frame moving at the instantaneous velocity of the Earth at the time of the
event, and this is the reference frame that $\mu_{\rm rel,G}$ is determined in.
The green histogram in Figure~\ref{fig-pm_comp} shows the distribution
of $\mu_{\rm rel,G}$ from our MCMC light curve modeling calculations
using both the Stanek and OGLE-pipeline reductions of the MDM data.
The spread in $\mu_{\rm rel,G}$ values is primarily due to the uncertainty
in the source trajectory angle, $\theta$, as discussed in Section~\ref{sec-lc}.

In Section~\ref{sec-hst_data}, we will present the relative lens-source proper
motion measurement from the HST observations. This measurement is made
with respect to the average motion of the Earth during the 6.4678 years between
the event and the HST observations.
If we assume that the HST observations are made in a Heliocentric
frame, the maximum error in the lens-source displacement is twice the
relative lens-source relative parallax or
$2\pi_{\rm rel} = 2{\rm AU}(1/D_L - 1/D_S) \simeq 0.26$ (assuming our
final result), which compares to our lens-source displacement measurement 
error of $1.3\,$mas, so the assumption of a Heliocentric reference frame is
a reasonable approximation. (The lens-source relative parallax is given
by $\pi_{\rm rel} = {\rm AU}\left(D_L^{-1}-D_S^{-1}\right)$. )
The HST measurements
also determine the direction of the lens-source relative proper motion, so they
determine the 2-dimensional relative proper motion, $\mubold_{\rm rel,H}$.

\section{HST Astrometry and Photometry}
\label{sec-hst_data}

We observed the OGLE-2005-BLG-169 source and lens stars for two HST
orbits as a part of HST Program GO-12541. On 2011 October 19, we
obtained images in three passbands, F814W, F555W, and F438W, using the 
Wide Field Camera 3-Ultraviolet-Visible (WFC3-UV) instrument.
We obtained $7\times 85\,$sec dithered F814W exposures, 
$8\times 175\,$sec dithered F555W exposures, and 
$6\times 349\,$sec dithered F438W exposures. Due to the
relatively short exposure times, we have been forced to limit
the amount of data read out for the F555W and F814W exposures.
Only 1k$\times$1k region of the CCDs were read out for these
passbands.

\begin{figure}
\epsscale{1.0}
\plotone{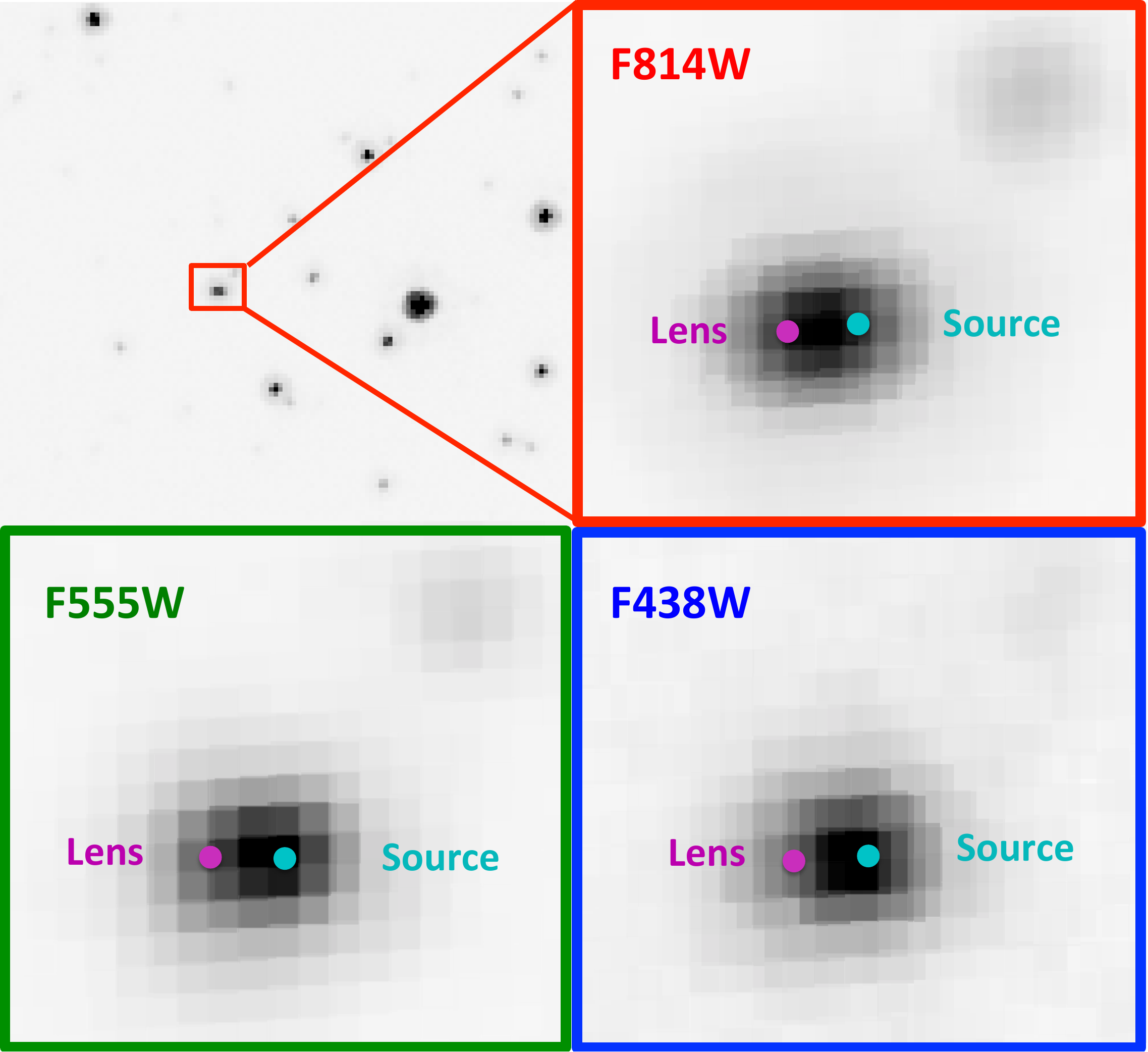}
\caption{The top-left panel shows a $4.9^{\prime\prime}\times 4.6^{\prime\prime}$
section of the stacked F814W images containing the target, and the top-right
panel shows a close-up of a $0.42^{\prime\prime}\times 0.37^{\prime\prime}$
region containing the lens and source stars. The lower-left and lower-right
panels show the same region using the stacked images in the F555W and
F438W passbands. The magenta spots on the left are the best fit lens locations
and the cyan spots on the right are the best fit source positions. These close-up
images are sums of all the dithered images with $100\times$ oversampling.
The distortion of the WFC3 images has been removed, and this results
in pixels shaped like parallelograms rather than squares.
\label{fig-HSTim}}
\end{figure}

The data were reduced following the method of \citet{andking00,andking04}. 
The dithered exposures are used to construct
an effective PSF from stars of a similar color to the target (\ie\ the
blended image of the source plus lens stars). Then, this effective
PSF is used to fit two stellar profiles to the blended target image.
The top-right and bottom two panels of Figure~\ref{fig-HSTim} show
close-ups of the blended source plus lens stars in the three
passbands, F814W, F555W, and F438W, which are the HST versions
of the $I$, $V$, and $B$-bands. The best fit locations of the 
lens and source stars are also indicated. In the F814W images,
both the stars have a brightness consistent with $I$-band source
brightness determined from light curve modeling, so there would
be ambiguity in the lens and source star identifications if we had
data in this passband alone. Fortunately, the lens is considerably
fainter than the source in the F555W and F438W passbands, and this
allows us to uniquely identify the lens and source stars. (The lens
is closer than the main sequence source, so it must be redder than
the source if it has the same magnitude in the $I$-band.)

The separation between
the lens and source stars is due to the $\Delta t = 6.4678\,$yr
interval between the event and the HST observations, so we can
determine the relative proper motion in the Heliocentric frame
by $\mubold_{\rm rel,H} = \mathbold{\Delta x}/\Delta t$,
where $ \mathbold{\Delta x}$ is the two-dimensional separation
between the lens and source stars as measured in the HST images.
In the Galactic coordinate system, we find
\begin{eqnarray}
\label{eq-pmI}
\mubold_{\rm rel,H}(l,b) &= (7.52\pm 0.27, 1.07\pm 0.28)\,{\rm mas/yr}\ \   ({\rm F814W}) \ ,\\
\label{eq-pmV}
\mubold_{\rm rel,H}(l,b) &= (7.17\pm 0.33, 1.88\pm 0.41)\,{\rm mas/yr}\ \   ({\rm F555W}) \ ,\\
\label{eq-pmB}
\mubold_{\rm rel,H}(l,b) &= (7.32\pm 0.67, 1.40\pm 0.82)\,{\rm mas/yr}\ \   ({\rm F438W}) \ .
\end{eqnarray}
The error bars for these $\mubold_{\rm rel,H}$ values are determined from
the dual star fits. First the fit $\chi^2$ values are renormalized to give 
$\chi^2/{\rm d.o.f.} = 1$. The original $\chi^2/{\rm d.o.f.}$ values were 1.25, 1.64, and 1.39
for the F814W, F555W, and F438W passbands, respectively. Then, we add $0.1\,$mas/yr
in quadrature and multiply the error bars by 1.5. These adjustments are meant to 
account for systematic uncertainties in the PSF models, and they
ensure that values from the different passbands are consistent. (The systematic errors
could probably be reduced by constructing different PSF models for the lens and
source stars rather than one PSF based on their average color.) Note that
the F438W and the Keck $H$-band \citep{batista15} proper motion values both fall in between the
F814W and F555W values, so there is no trend with color.
We combine the best two measurements from 
equations~\ref{eq-pmI}-\ref{eq-pmB} (F814W and F555W),
to obtain the measurement we will use as our final measurement of the
lens-source relative proper motion
\begin{equation}
\label{eq-pmVI}
\mubold_{\rm rel,H}(l,b) = (7.39\pm 0.20, 1.33\pm 0.23)\,{\rm mas/yr} \ .
\end{equation}

The direction of proper motion is about $\sim 10^\circ$ from the Galactic longitude,
$l$, direction. This is expected for a Galactic disk lens about half way to the 
center of the Galaxy. Due to the Galaxy's flat rotation curve, we and the lens
system move at about the same velocity, but the source star in the Galactic
bulge doesn't share this rotation, so the relative lens-source proper motion is
typically $\mu_{\rm rel} \approx 220\,{\rm km/s}/8.3\,{\rm kpc} = 5.6\,$mas
in the direction of Galactic rotation. The velocity dispersion is dominated 
by the Galactic bulge one-dimensional velocity dispersion of 
$\sim 100\,$km/s, which implies a one-dimensional proper motion dispersion of 
$2.5\,$mas/yr. So typically, the relative proper motion of lens half way to the
bulge should be within $\sim 30^\circ$ of the Galactic rotation direction, while
an event like OGLE-2005-BLG-169, with a higher than average $\mu_{\rm rel}$, 
would typically have a relative proper motion within $\sim 30^\circ$ of the 
Galactic rotation direction.

These fits also return the magnitudes of the source and lens stars. To put these
on a standard scale, we calibrate the $V$ (F555W) and $I$ (F814W) to the 
OGLE-III catalog \citep{ogle3-phot}. We find 12 uncrowded stars with 
$ 0.86 < V-I < 2.24$ and $I < 19$ that we use for this calibration. The
scatter in these calibrations are about $1\,$\%, so the formal uncertainty
in the calibration is 0.3-$0.4\,$\%. 

With this calibration, the best fit $V$ and $I$ source magnitudes are
\begin{eqnarray}
\label{eq-Smags}
V_S =& 22.212 \pm 0.041 \ \ \ \ I_S = 20.555 \pm 0.054 \\
\label{eq-Lmags}
V_L =& 22.783 \pm 0.067  \ \ \ \ I_L = 20.493 \pm 0.051 \\
\label{eq-totmags}
V_{\rm tot} =& 21.704 \pm 0.020 \ \ \ \ \ I_{\rm tot} = 19.771 \pm 0.020 ,
\end{eqnarray}
where $V_{\rm tot} $ and $I_{\rm tot}$ refer to the magnitudes corresponding
to the combined brightness of the lens plus source stars. The formal 
errors on $V_{\rm tot}$ and $I_{\rm tot}$ are actually only about $0.003\,$mag,
but we use 0.020 mag to account for calibration uncertainties. The uncertainties
on the source and lens magnitudes are significantly larger than the uncertainty
on the combined lens+source magnitude. This is due to the fact that the lens
and source are not fully resolved, which allows correlated uncertainties
where the lens and source can trade flux with slight modifications in their
best fit positions.

The F438W data were calibrated to the Vega-magnitude scale by comparison
to a reduction of the same data using DolPHOT, which is an updated version
of HSTphot \citep{hstphot} by the same author. This gives
$B_S = 23.82 \pm 0.06$ and $B_L = 24.74 \pm 0.15$.

\subsection{Confirmation of Planetary Signal Prediction}
\label{sec-conf}

Our HST measurements of the lens-source relative proper motion, 
$\mubold_{\rm rel,H}$, are made in a reference frame that is 
indistinguishable from the Heliocentric reference frame, but the light
curve measurements are made in the Geocentric reference frame
that moves with the Earth at the time of the event. These reference
frames differ by the velocity of the Earth at the time of the event 
projected onto the plane of the sky at the time of the event. This 
projected velocity is
\begin{equation}
\label{eq-vearth}
(v_{\oplus N}, v_{\oplus E}) = (3.15, 18.51)\,{\rm km/sec} = (0.665, 3.905)\,{\rm AU/yr}\ ,
\end{equation}
and the relationship between the geocentric and heliocentric relative proper 
motions is
\begin{equation}
\label{eq-murelHG}
\mubold_{\rm rel,H} = \mubold_{\rm rel,G} + {\pi_{\rm rel}\over {\rm AU}} \vbold_\oplus \ .
\end{equation}
Converting to Galactic coordinates, we have
$(v_{\oplus l}, v_{\oplus b}) = (3.74, -1.30)\,$AU/yr, so equation~\ref{eq-murelHG} 
becomes
\begin{equation}
\label{eq-murelGH1}
\mubold_{\rm rel,G}(l,b) = \mubold_{\rm rel,H}(l,b) + {\pi_{\rm rel}\over {\rm yr}} (-3.74, 1,30) \\
\end{equation}
in Galactic coordinates or
\begin{equation}
\label{eq-murelGH2}
\mubold_{\rm rel,G}(l,b)  = (7.39, 1.33) + {\pi_{\rm rel}\over 0.13\,{\rm mas}} (-0.49, 0.17) \ ,
\end{equation}
after substituting our measured value from equation~\ref{eq-pmVI}. We choose 
$\pi_{\rm rel} = 0.13\,$mas as our reference value in equation~\ref{eq-murelGH2}
because this is a round number close to our best fit final value. It corresponds to
a lens system at a distance of about $4\,$kpc, about half-way to the Galactic
center.

However, we'd like to compare our measurement of $\mubold_{\rm rel,H}$ to
the light curve prediction of $\mu_{\rm rel,G}$ with no reference to our best
fit lens distance, in order to have
a relatively pure test of the relative proper motion
prediction from the light curve.
We therefore convert the measured Heliocentric relative proper motion measurement 
to a probability distribution for $\mu_{\rm rel,G}$ using a Bayesian analysis with
the Galactic model of \citet{bennett14}. This analysis implicitly assumes that potential
primary lens mass for the OGLE-2005-BLG-169 event is equally likely to host a planet
with the measured mass ratio, but it makes no assumptions about the location of
the lens system or source star.

The red cross-hatched histogram in Figure~\ref{fig-pm_comp} indicates the 
distribution of $\mu_{\rm rel,G}$ values consistent with the HST $\mubold_{\rm rel,H}$
measurement. The HST values for $\mu_{\rm rel,G}$ are near the extreme low-$\mu_{\rm rel,G}$
edge of the light curve distribution, shown in green. However, the histograms cross at
a probability of about 80\% of the maximum of each curve. Thus, the HST measurement
is clearly consistent with the $\mu_{\rm rel,G}$ predictions from the light curve.

The light curve prediction of $\mu_{\rm rel,G}$ comes directly from the planetary 
light curve feature, because this is the only light curve feature that resolves the
finite source size and determines the source radius crossing time, $t_*$. Thus,
our HST confirmation of the $\mu_{\rm rel,G}$ is also a confirmation of the
planetary interpretation of the OGLE-2005-BLG-169 light curve. This is the first
such confirmation for a planetary microlensing event.

\subsection{Comparison to Keck Adaptive Optics Measurements}
\label{sec-Keck}

In July, 2013, 
a subset of us obtained 15 Keck NIRC2 Adaptive optics $H$-band images with seeing of 
$\sim 55\,$mas. These high resolution images resolved the lens and source stars, so that
their separation could be measured, some $8.2121\,$years after the microlensing
event peak. This allowed an independent measurement of the lens-source relative
proper motion \citep{batista15},
\begin{equation}
\label{eq-pmKeck}
\mubold_{\rm rel,H}(l,b)[{\rm Keck}] = (7.28\pm 0.12, 1.54\pm 0.12)\,{\rm mas/yr} \ .
\end{equation}
This measurement is obviously quite consistent with our HST measurement given
above (equation~\ref{eq-pmVI}) as both the $l$ and $b$ components are within
1-$\sigma$ of our values.

Both of these $\mubold_{\rm rel,H}$ measurements make the assumption that the
lens and source are coincident during the microlensing event, but we can also use
the $1.7443\,$yr interval between the HST and Keck observations to work out
the separation at the time of the event between the stars we identify as the 
lens and source. This gives a separation of
\begin{equation}
\label{eq-dldb}
(\Delta l, \Delta b) = (3.5 \pm 7.2, -6.4 \pm 7.8)\,{\rm mas} \ ,
\end{equation}
between these lens and source stars at the time of the event. These are
consistent at $< 1\,\sigma$ with our identification of the lens and source
stars, whose separation was $\theta_E u_{\rm min} = 0.0011\,$mas at the
time of the event.

This measurement is also sufficient to rule out
the possibility that the detected flux is due to a binary companion to the 
lens. A possible binary companion to the lens that orbits within $30\,$mas 
of the primary lens star would strongly perturb the light curve, so such
a close companion is excluded by the light curve observations, which see no
evidence of such a companion \citep{bennett07,gould-1dpar}. Thus, this
light curve constraint combined with the combined Keck+HST measurements,
excludes the possibility that a binary companion to the lens is responsible for the
flux that we attribute to the lens stars.

\subsection{Lens System Properties from HST Measurements}
\label{sec-const}

As discussed in Section~\ref{sec-radius}, the angular Einstein radius, 
$\theta_E = \theta_* t_E/t_*$, can be determined from light curve parameters,
as long as the angular source size, $\theta_*$, can be determined from the 
source brightness and color. The determination of $\theta_E$ allows us to use
the following relation \citep{bennett_rev,gaudi_araa}
\begin{equation}
M_L = {c^2\over 4G} \theta_E^2 {D_S D_L\over D_S - D_L} 
       =  {c^2\over 4G} \theta_E^2 {{\rm AU}\over \pi_{\rm rel}}
       = 0.9823\,\msun \left({\theta_E\over 1\,{\rm mas}}\right)^2\left({x\over 1-x}\right)
       \left({D_S\over 8\,{\rm kpc}}\right) \ ,
\label{eq-m_thetaE}
\end{equation}
where $x = D_L/D_S$. This expression
can be considered to be a mass-distance relation, since $D_S$ is approximately
known. As can be seen from 
Table~\ref{tab-mparams}, the light curve does not determine $\theta_E$ very 
precisely, due to the correlated uncertainty in $\theta$ and $t_*$. However, as
Figure~\ref{fig-pm_comp} indicates, the HST observations rule out a large fraction of the
$\mu_{\rm rel,G}$ values that are compatible with the light curve. Since 
$\theta_E = \mu_{\rm rel,G} t_E$, this implies that much of the $\theta_E$ range
allowed by the light curve is now excluded by the HST data. This, in turn,
has an effect on other parameters, such as the planet:star mass ratio,
which is $q \approx 6\times 10^{-5}$ for the $\theta \approx 1.6$ solutions,
compared to $q \approx 8\times 10^{-5}$ for the $\theta \approx 1.0$
solutions. So, the HST data drive the planetary mass fraction to a somewhat lower
value.

\begin{deluxetable}{cccc}
\tablecaption{Physical Parameters\label{tab-pparam}}
\tablewidth{0pt}
\tablehead{
\colhead{Parameter}  & \colhead{units} & \colhead{value} & \colhead{2-$\sigma$ range} }
\startdata 
$D_L $ & kpc & $4.1\pm 0.4$ & 3.3-4.8 \\
$M_\star$ & $\msun$ & $0.69\pm 0.02$ & 0.64-0.73 \\
$m_p$ & $\mearth$ & $14.1\pm 0.9$ & 12.4-15.9 \\
$a_\perp$ & AU & $3.5\pm 0.3$ & 2.9-4.0  \\
$a_{3d}$ & AU & $4.0{+2.2\atop -0.6}$ & 3.0-14.0  \\
\enddata
\tablecomments{ Uncertainties are
1-$\sigma$ parameter ranges. }
\end{deluxetable}

To solve for the planetary system parameters, we sum over our MCMC results
as in Subsection~\ref{sec-conf} with the weighting by the Galactic model parameters
consistent with the HST $\mubold_{\rm rel,H}$ measurement, but now we add
the HST lens brightness constraints, as well. In this sum, we randomly
select source and lens distances that are consistent with the 
mass-distance relation (equation~\ref{eq-m_thetaE}). In order to check this
consistency, we must invoke a mass-luminosity relation. We use the mass-luminsity
relations of \citet{henry93}, \citet{henry99} and \citet{delfosse00}.
For $M_L > 0.66\,\msun$, we use the \citet{henry93} relation; for
$0.12\,\msun < M_L < 0.54\,\msun$, we use the \citet{delfosse00} relation; and for
$0.07 \,\msun < M_L < 0.10\,\msun$, we use the \citet{henry99} relation. In between these
mass ranges, we linearly interpolate between the two relations used on the
boundaries. That is we interpolate between the \citet{henry93} and the \citet{delfosse00}
relations for $0.54\,\msun < M_L < 0.66\,\msun$, and we interpolate between the
\citet{delfosse00} and \citet{henry99} relations for $0.10\,\msun < M_L < 0.12\,\msun$.

At a Galactic latitude of $ b = -4.7402^\circ$, and a lens distance of $\sim 4\,$kpc, the lens system
is likely to be behind most, but not all, of the dust that is in the foreground of the source. 
We assume a dust scale height of $h_{\rm dust} = 0.10\pm 0.02$kpc, so that the
extinction in the foreground of the lens is given by
\begin{equation}
A_{i,L} = {1-e^{-|D_L/(h_{\rm dust} \sin b)|}\over 1-e^{-|D_S/(h_{\rm dust}\sin b)|}} A_{i,S} \ ,
\end{equation}
where the index $i$ refers to the passband: $V$, $I$, or $H$. For each model in the
Markov Chain, the $h_{\rm dust}$ value is selected randomly from a Gaussian distribution.
We assume error bars of $\sigma_V = 0.10$ and $\sigma_I = 0.07$ magnitudes for the
combined uncertainty in the mass-luminosity relations and the lens star extinction
estimate. The results of this final sum over the Markov Chain are given in 
Table~\ref{tab-pparam}. The host star is a $M_\star = 0.69\pm 0.02\,\msun$ K-dwarf, orbited
by a planet of about Uranus' mass at $m_p = 14.1\pm 0.9\,\mearth$, at a projected
separation of $a_\perp = 3.5\pm 0.3\,$AU. Assuming a random orientation, this
implies 3-dimensional separation of $a_{3d} = 4.0{+2.2\atop -0.6}\,$AU. This
planet then has the mass of an ice-giant in a Jupiter-like orbit at about
twice the nominal snow-line distance of $2.7(M/\msun)\,{\rm AU} \simeq 1.9\,$AU.
This is similar to a number of other planets found by microlensing
\citep{ogle390,sumi10,muraki11,moa328}, which can be interpreted as
examples of ``failed Jupiter cores\rlap." These would be planets that grew
by accumulation of solids, as Jupiter's core is thought to have done
\citep{lissauer_araa}, if the core accretion model is correct. These 
``failed Jupiter core" planets are thought to be common around the low-mass
stars probed by the microlensing exoplanet search method.

\section{Discussion and Conclusions}
\label{sec-conclude}

We report the first detection of an exoplanet microlens host star found at
the separation predicted by the exoplanet feature in the microlensing light
curve. Together with a companion paper based on Keck data \citep{batista15},
this provides the first confirmation of a microlensing planetary signal, in the
sense that the planetary interpretation of a light curve feature predicted
the lens-source relative proper motion, which we, and \citet{batista15} have confirmed.

The resulting system is a Uranus-mass planet orbiting a K-dwarf at about twice
the snow-line, which fits the properties of a ``failed Jupiter core" planet,
predicted by core accretion \citep{laughlin04}.

This is also the first demonstration of the primary exoplanet host star mass
measurement method \citep{bennett07} planned for WFIRST \citep{WFIRST_AFTA}
and EUCLID \citep{penny13}.
While, the host star mass might plausibly be inferred from just the brightness of the
lens star \citep{bennett06,dong-ogle71,batista14}, but to be highly confident that
the measured star is actually the lens star \citep{janczak10}, it is necessary to measure the 
lens-source relative proper motion and show that it is consistent with the 
prediction from the light curve. This will be even more important for the 
WFIRST exoplanet microlensing survey, because it will work in more crowded fields,
where the microlensing rate is highest.

The HST data presented her provide an extremely high S/N measurement of the
lens-source relative proper motion. The lens-source separation is measured at
28-$\sigma$ in the F814W band, 22-$\sigma$ in the F555W band, and 11-$\sigma$
in the F438W band. This is partly because this event is a favorable one for
such measurement \citep{henderson14}, but also because it took a while for
the HST TAC to recognize the importance of such measurements. As discussed in
\citet{bennett07}, this measurement could easily have been made 4 years earlier.

It was not necessary to achieve the photon noise limit in our HST astrometry
measurements because of the high S/N in the HST data. As the
discussion in Section~\ref{sec-hst_data} indicates, 
our error bars are probably about a factor of two above the photon noise limit.
There are several things that can be done to improve the analysis. One improvement
would be to add a second iteration of PSF fitting to determine the source and
lens properties. The first iteration determines the approximate source and lens
colors, but the stars selected to make the PSF models are matched to the
average lens+source color. In a second iteration, new PSF models could be make
to match the lens and source star, and second round of fitting could be done with
custom PSF models for the lens and source stars.

An additional improvement in the method would be to fit more than two sources
in the HST images in the vicinity of the target star. In the close-ups in the
top-right and bottom two panels of Figure~\ref{fig-HSTim}, there is a faint star
in the upper right corner. If this star were brighter or if the lens or source were
much fainter, the PSF wings of this star could interfere with the lens and/or 
source star fits. The solution is then to also fit for that star. We expect to use
this method for other targets that have been observed in Program GO-12541.

\acknowledgments 
Based on observations made with the NASA/ESA Hubble Space Telescope, 
obtained at the Space Telescope Science Institute (STScI), which is operated by the 
Association of Universities for Research in Astronomy, Inc., under NASA contract 
NAS 5-26555. These observations are associated with programs \# 12541 and 13417.
D.P.B.,A.B., and D.S.\  were supported by NASA through grants from the STScI
and grant NASA-NNX12AF54G.
A.G.\ and B.S.G.\ were supported by NSF grant AST 110347 and
by NASA grant NNX12AB99G.
S.D. is supported by Òthe Strategic Priority Research Program- The
Emergence of Cosmological StructuresÓ of the Chinese Academy of
Sciences (grant No. XDB09000000).
The OGLE project has received funding from the European Research Council
under the European Community's Seventh Framework Programme
(FP7/2007-2013) / ERC grant agreement no. 246678 to AU.

\end{document}